\documentclass[twocolumn,showpacs,prb]{revtex4}
\usepackage{epsfig}

\newcommand{\be}{\begin{equation}}
\newcommand{\ee}{\end{equation}}
\newcommand{\beqn}{\begin{eqnarray}}
\newcommand{\eeqn}{\end{eqnarray}}
\newcommand{\nsw}{N_{\mathrm{sweep}}}
\newcommand{\nsa}{N_{\mathrm{samp}}}
\newcommand{\varql}{{\mathrm{Var}}(\ql)}
\newcommand{\ql}{q_{l}}

\begin{document}

\title{Monte Carlo simulations of spin glasses at low temperatures:
Effects of free boundary conditions}

\author{Helmut G. Katzgraber}
\email{dummkopf@physics.ucsc.edu}
\altaffiliation{Present address: Department of Physics, University of
California, Davis, CA 95616}

\author{A. P. Young}
\email{peter@bartok.ucsc.edu}
\homepage{http://bartok.ucsc.edu/peter}
\altaffiliation{Present address:
Department of Theoretical Physics, 1, Keble Road, Oxford OX1 3NP,
England}
\affiliation{Department of Physics,
University of California,
Santa Cruz, California 95064}

\date{\today}

\begin{abstract}
We present results of Monte Carlo simulations, using parallel tempering,
on the three- and four-dimensional Edwards-Anderson
Ising spin glass with Gaussian couplings at
low temperatures with free boundary conditions.  Our results suggest that
the surface of large-scale low-energy excitations may be space filling. 
The data implies that
the energy of these excitations increases with increasing system size for
small systems, but we see evidence in three dimensions, where we have a
greater range of sizes, for a crossover to a regime where the energy 
is independent of system size, in accordance with replica
symmetry breaking.
\end{abstract}

\maketitle

\section{Introduction}
\label{introduction}

Recently there have been several
studies\cite{palassini:99,krzakala:00,palassini:00,%
marinari:00,marinari:00b,middleton:99,middleton:01}
that attempt
to better understand
the nature of the spin-glass state through zero-temperature calculations.
Whereas these calculations have used different versions of the {\em heuristic}\/
genetic algorithm, recently the {\em exact}\/ branch and cut algorithm has
been applied \cite{palassini:01a} to the Edwards-Anderson
Ising spin glass with {\em free}\/ boundary conditions. In part, this choice
of boundary conditions has been motivated by the algorithm being more efficient for
free than periodic boundary conditions. However, in addition,
use of a different boundary condition
also allows one to study the role that boundary conditions play, and perhaps
eventually to deduce the optimal choice
of boundary conditions for numerical studies.

There have been two principle scenarios proposed for the spin-glass state.
In replica symmetry breaking (RSB)\cite{parisi:79,mezard:87} 
large-scale low-energy excitations of the system cost a 
finite energy in the thermodynamic limit 
and have a surface that is space filling, i.e.~the fractal
dimension of the surface, $d_s$, is equal to the space dimension $d$.
In the other commonly discussed scenario, called the
droplet picture (DP),\cite{fisher:87,huse:87,fisher:88}
it is argued that the lowest-energy excitation
with linear spatial extent $L$ and
involving a given spin
typically 
costs an energy $L^\theta$, where $\theta$ is a (positive) exponent.
Hence, in the thermodynamic limit, these excitations cost an infinite
energy. In addition, it is predicted that the surface of these excitations
is fractal with $d_s < d$.

An intermediate picture between RSB and the DP has also been
proposed by Krzakala and Martin\cite{krzakala:00} and Palassini and
Young\cite{palassini:00} (KMPY) on the basis of
numerical calculations at
$T=0$ (see also Ref.~\onlinecite{katzgraber:01} for analogous calculations at
finite-$T$). In this scenario,
large-scale low-energy excitations
cost a finite energy in the thermodynamic limit but their surface is fractal.
There are
two exponents\cite{krzakala:00,palassini:00} that describe the 
energy dependence of the system
size: $\theta \ (> 0)$, where $L^\theta$ is the typical change
in energy when the boundary conditions are changed, as originally proposed by
the droplet model, and $\theta^\prime$, where $L^{\theta^\prime}$ 
characterizes the energy of system-size excitations
thermally excited within the system for a {\em fixed} set of boundary
conditions.

The data of Refs.~\onlinecite{krzakala:00} and \onlinecite{palassini:00} 
for periodic boundary
conditions agree well with the KMPY scenario, while agreement with RSB would
require large corrections to scaling. However,
the results of Ref.~\onlinecite{palassini:01a}
appear to be somewhat different. They are compatible with RSB but it cannot be
excluded that this is a finite-size effect due to large corrections with
free boundary conditions, so the KMPY or droplet pictures cannot be
ruled out.

In this work we perform finite temperature
Monte Carlo simulations of the three- and
four-dimensional Edwards-Anderson (EA) Ising spin glass with free boundary
conditions at finite but low temperatures.
We find results that are consistent with the
ground-state calculations of Ref.~\onlinecite{palassini:01a}.

To investigate differences between the models mentioned it is useful
to look at the distribution of the 
spin overlap,\cite{marinari:00a,reger:90,marinari:98a,marinari:99} $P(q)$.
In the droplet picture,\cite{bray:86,moore:98,bokil:00}
$P(q)$ is trivial in the thermodynamic limit,
i.e., there
are only two peaks at $\pm q_{\rm EA}$
($q_{\rm EA}$ is
the Edwards-Anderson order parameter). For finite systems of linear size $L$, 
there is a tail with weight $\sim L^{-\theta}$ down to $q = 0$.
Conversely, RSB predicts a tail with a finite weight down to $q = 0$
independent of system size. In addition, the variance of the distribution of
the link overlap $q_l$ (introduced below) can shed some light on the 
surface of the excitations: the droplet picture predicts that the variance of the
link overlap has a power-law decay $\varql \sim L^{-\mu_l}$,  
where\cite{katzgraber:01} $\mu_l =
\theta^\prime +2(d - d_s)$. Because RSB predicts that the surface of the
low-energy excitations is space-filling and, in addition, that system-size
excitations cost only a finite energy, one expects that $\varql \to {\rm
const}$ for $L \to \infty$.
 
The layout of the paper is as follows: In Sec.~\ref{model-observables},
we describe the model as well as the observables measured, while in
Sec.~\ref{equilibration}, we discuss our equilibration tests for the parallel
tempering Monte Carlo method. Our results are presented 
in Sec.~\ref{results} and the conclusions summarized in
Sec.~\ref{conclusions}.

\section{Model and Observables}
\label{model-observables}

\begin{figure}
\centerline{\epsfxsize=5cm \epsfbox{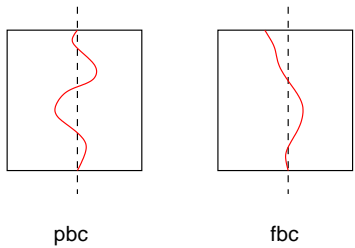}}
\caption{
Sketch of the effect of different boundary conditions on domain walls. For 
simplicity, we show a sample in two dimensions. The solid line with curvature
represents a domain wall. Periodic boundary conditions (pbc)
restrict the domain wall to enter and exit the sample at corresponding
points on the top and bottom surfaces. This is not the case for free 
boundary conditions (fbc) where the domain wall can enter and exit the
sample at arbitrary points.
}
\label{fbc.sketch}
\end{figure}

The Hamiltonian of the Edwards-Anderson Ising spin glass is given by
\begin{equation}
{\cal H} = -\sum_{\langle i,j \rangle} J_{ij} S_i S_j ,
\label{ham}
\end{equation}
where the sites $i$ lie on a hypercubic lattice in dimension 
$d=3$ or 4 with $N=L^d$ sites [$L \le 8$ in three dimensions (3D), $L \le 5$ in 
four dimensions (4D)], 
$S_i=\pm 1$, and the $J_{ij}$ are the nearest-neighbor interactions chosen 
according to a Gaussian distribution with zero mean and standard deviation 
unity. Free boundary conditions are applied.
Applying free 
boundary conditions has the advantage that domain walls are not restricted 
to enter and exit the sample at the corresponding point on opposite sides 
of the system, as sketched in Fig.~\ref{fbc.sketch}.

Our attention focuses primarily on two quantities: 
the spin overlap $q$, defined by
\begin{equation}
q = {1 \over N} \sum_{i=1}^N S_i^{(1)} S_i^{(2)} ,
\label{q}
\end{equation}
where ``$(1)$'' and ``$(2)$'' refer to two copies (replicas) of 
the system with identical bonds, and the link overlap, $\ql$, defined by
\begin{equation}
\ql = {1 \over N_b} \sum_{\langle i, j \rangle}
S_i^{(1)} S_j^{(2)} 
S_i^{(2)} S_j^{(2)} .
\label{ql}
\end{equation}
In the last equation, $N_b = dL^{d-1}(L - 1)$ is the number of bonds,
and the sum is over all pairs of spins $i$ and $j$ 
connected by bonds. 

If two spin configurations differ by flipping a large cluster then $q$
differs from unity by an amount proportional to the {\em volume}\/ of the
cluster while $\ql$ differs from unity by an amount proportional to the {\em
surface}\/ of the cluster.

\section{Equilibration}
\label{equilibration}

For the simulations, we use the parallel tempering Monte Carlo
method\cite{hukushima:96,marinari:98b} as it allows us to
study larger systems at lower temperatures.
We test equilibration with the method introduced by Katzgraber {\em et al.}~in 
Ref.~\onlinecite{katzgraber:01} for short-range spin glasses with 
a Gaussian distribution of exchange interactions. It depends on an
expression that relates the average energy per spin $|U|$ 
to the average link overlap:
\be
[\,\langle \ql \rangle \, ]_{\rm av}  =  1 - {T |U| N\over N_b},
\label{equilrel}
\ee
where $[\cdots ]_{\rm av}$ denotes an average over samples, and
$\langle \cdots \rangle$ denotes a thermal average.

\begin{figure}
\centerline{\epsfxsize=\columnwidth \epsfbox{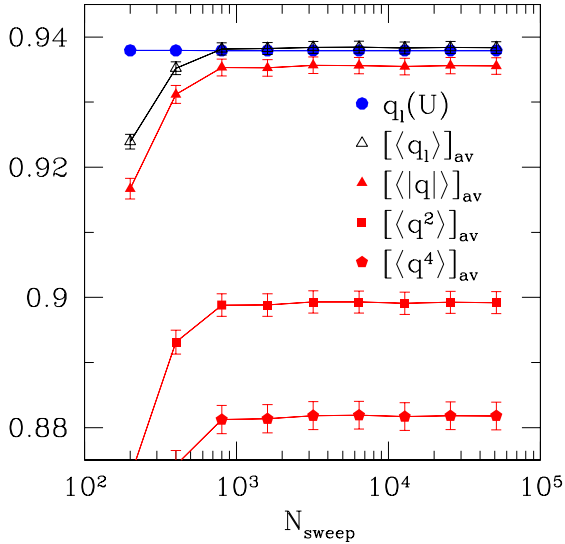}}
\caption{
The average link overlap $[\,\langle \ql\rangle\,]_{\rm av}$,
and $\ql(U)$ [the right-hand side of
Eq.~(\ref{equilrel})], as a 
function of the number of Monte Carlo sweeps, $\nsw$, which 
each of the replicas perform for $d=3$, $T=0.2$, and $L=4$.
Thermal averaging was performed 
over the last half of the sweeps indicated.
The two sets of data approach 
each other from opposite directions and then do not appear to 
change at larger number of sweeps, indicating that they have 
equilibrated. We also show data for the average first, second and fourth 
moments of $q$. They appear to be independent of the number of 
sweeps once the data for $\ql$ has equilibrated.
}
\label{equil}
\end{figure}

We choose a set of temperatures $T_i, i = 1, 2, \cdots , N_T$, so 
that the acceptance ratios for the global moves are satisfactory, 
typically greater than $0.3$ for $d = 3, 4$.
The simulation is started with 
randomly chosen spins so that all replicas are uncorrelated. 
This has the effect that both sides of Eq.~(\ref{equilrel}) 
are approached from opposite directions. Once they agree,
the system is in equilibrium, as can be seen in Fig.~\ref{equil} for
$T = 0.2$, the lowest temperature simulated, with $L = 4$, $d = 3$. 
We show data for a smaller size as it allows us to generate more 
samples for longer equilibration times to better illustrate the method. 
For larger system sizes, we stop the simulation once the data for 
$[\,\langle\ql \rangle \,]_{\rm av}$ and $1-T|U|N/N_b$ agree.

In order to calculate the spin and link overlaps in Eqs.~(\ref{q}) and
(\ref{ql})
we use two replicas for each temperature, and so the total number of replicas
in the simulations is $2 N_T$.

\section{Results}
\label{results}

\subsection{Three dimensions}
\label{results.3d}

In Table~\ref{3d-tab-fbc}, we show $\nsa$, the
number of samples, $\nsw$, the total number of sweeps performed by each
set of spins (replicas), and $N_T$,
the number of temperature values, used in the simulations.
For all sizes the largest temperature is 2.0 and the lowest 0.20
(to be compared with $T_c \simeq 0.95$).\cite{marinari:98}
The temperatures are chosen in order for the acceptance ratios of the global
moves to be typically bigger than 0.3 for the largest size of$L = 8$.

\begin{table}
\caption{
\label{3d-tab-fbc}
Parameters of the simulations in three dimensions with
free boundary conditions. $\nsa$ is the
number of samples (i.e., sets of bonds), $\nsw$ is the total number of sweeps
simulated for each of the $2 N_T$ replicas for a single sample,
and $N_T$ is the number of
temperatures used in the parallel tempering method.
}
\begin{tabular*}{\columnwidth}{@{\extracolsep{\fill}} c r r l }
\hline
\hline
$L$  &  $\nsa$  & $\nsw$ & $N_T$  \\ 
\hline
3 & 20000  & $1.0 \times 10^4$  &   18 \\
4 & 20000  & $1.0 \times 10^4$  &   18 \\
5 & 10000  & $1.6 \times 10^5$  &   18 \\
6 & 10000  & $3.0 \times 10^5$  &   18 \\
7 &  5000  & $1.0 \times 10^6$  &   18 \\
8 &  5000  & $1.0 \times 10^6$  &   18 \\
9 &  5000  & $3.0 \times 10^6$  &   18 \\
\hline
\hline
\end{tabular*}
\end{table}

\begin{figure}
\centerline{\epsfxsize=\columnwidth \epsfbox{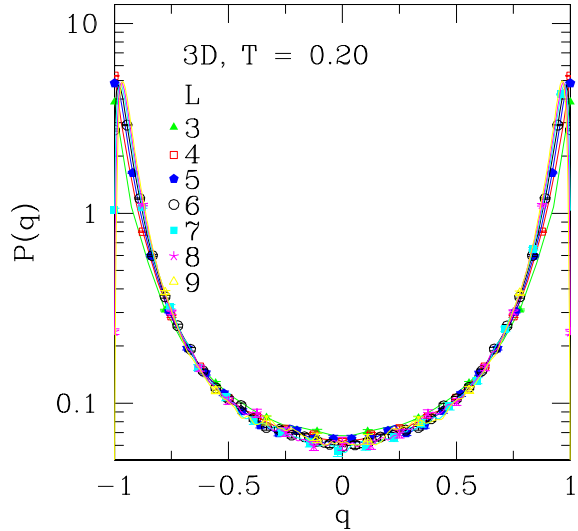}}
\caption{
Data for the overlap distribution $P(q)$ in 3D at $T=0.20$ with free
boundary conditions. Note
that the vertical scale is logarithmic to better make visible both
the peaks at large $q$ and the tail down to $q=0$.
The lines go
through all the data points but, for clarity, only some of the data points are
shown as points.
}
\label{pq0.20_3d-fbc}
\end{figure}

\begin{figure}
\centerline{\epsfxsize=\columnwidth \epsfbox{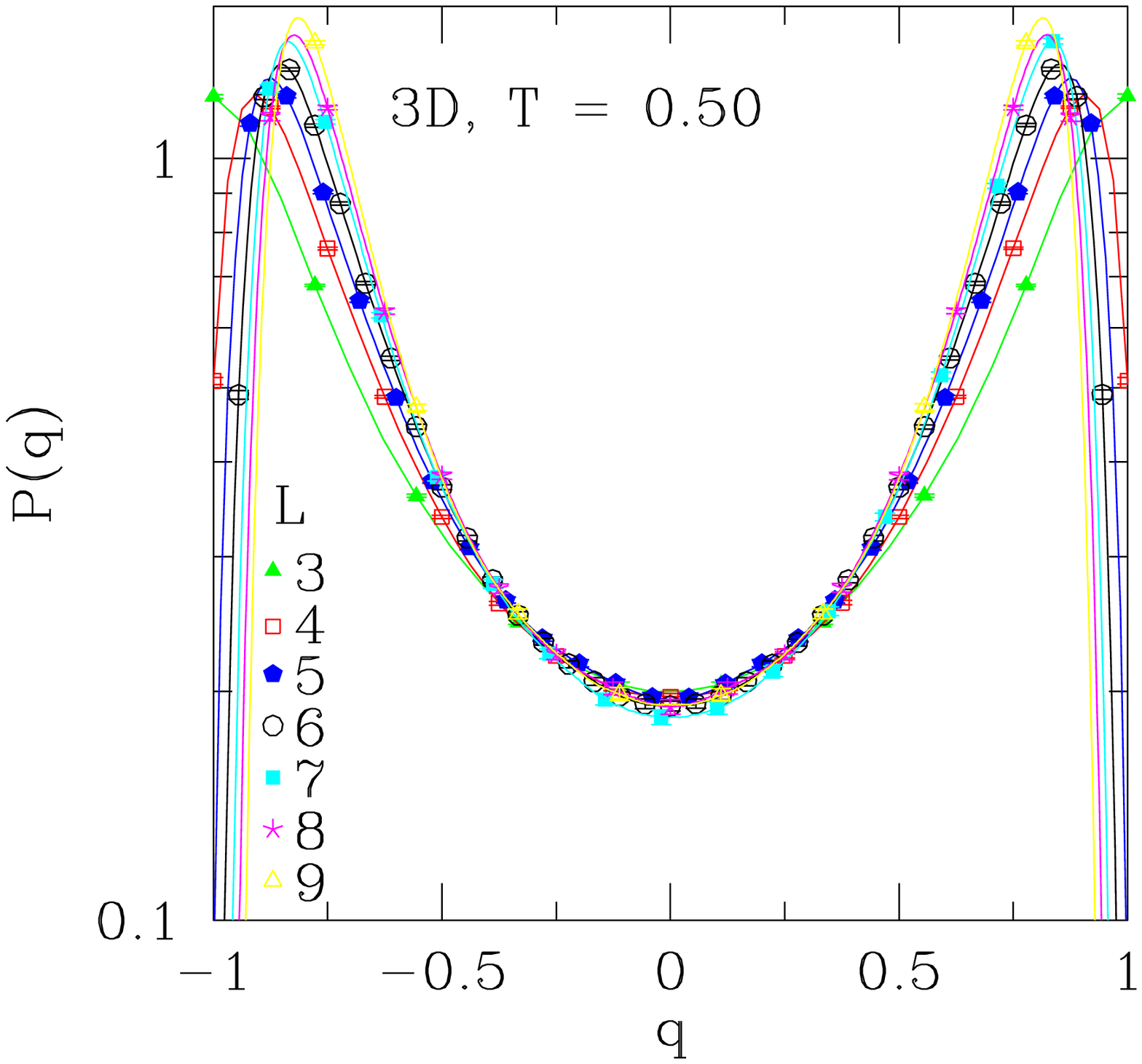}}
\caption{
Same as for Fig.~\ref{pq0.20_3d-fbc} but at $T=0.50$.
}
\label{pq0.50_3d-fbc}
\end{figure}

\begin{figure}
\centerline{\epsfxsize=\columnwidth \epsfbox{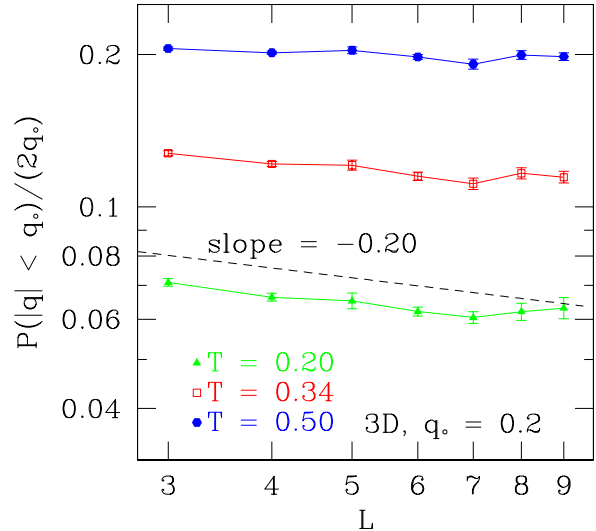}}
\caption{ Log-log plot of $P(0)$, the spin overlap at $q=0$, against $L$ in 
3D with free boundary conditions averaged over the range $|q| < q_\circ =
0.20$.  
The data slightly depend on system size. The dashed line has 
slope $-0.20$, the estimated value of $-\theta$ according to the
droplet picture. 
}
\label{p0_3d-fbc}
\end{figure}

Figures \ref{pq0.20_3d-fbc} and \ref{pq0.50_3d-fbc} show data for the spin
overlap distribution $P(q)$ for temperatures 0.20 and 0.50 respectively.
For low enough temperatures, our data are consistent with $P(0) \propto T$.
There is a large peak for large values of $q$ and a tail that 
depends slightly on the system size.
To determine more precisely the size dependence of $P(0)$, we average over
data points with $|q| < q_\circ$, where $q_\circ  = 0.20$. Different values for
$q_\circ$ give comparable results within error bars.
The results are shown
in Fig.~\ref{p0_3d-fbc}.
Since the droplet model predicts that $P(0)$ should vary as $L^{-\theta}$,
and the value of
$\theta$ obtained in numerical studies involving boundary condition
changes\cite{hartmann:99,bray:84,mcmillan:84} is about $0.20$, we indicate,
by the dashed line in  Fig.~\ref{p0_3d-fbc}, a slope of
$-0.20$. 
For low temperatures and small sizes the data are compatible
with  this droplet theory prediction.
However,  we see evidence for
crossover to a behavior where $P(0)$ is independent of $L$
at larger system sizes, which would be consistent with RSB.

Fitting the data in Fig.~\ref{p0_3d-fbc} to the
form $a L^{-\theta^\prime}$ for $L \le 7$, we obtain 
$\theta^\prime= 0.19 \pm 0.03$ for $T=0.20$, 
$\theta^\prime=0.15 \pm 0.02$ for $T=0.34$, and
$\theta^\prime=0.05 \pm 0.02$ for $T=0.50$. 
The goodness-of-fit probabilities $Q$ for these
fits\cite{press:95}
are $0.912$, $0.703$, and $0.384$ for $T = 0.20$, $0.34$, and $0.50$,
respectively. If we fix $\theta^\prime = 0$, the RSB prediction,
and attempt a fit to all values of $L$ as well as a subset with
$L \le 7$ the goodness-of-fit probabilities are smaller than $10^{-6}$
for $T = 0.20$ and $0.34$ and $0.017$ for $T = 0.50$ , which are 
rather poor. We also find that fixing $\theta^\prime = 0$ for $L \ge 7$ 
is more probable than $\theta^\prime = 0.20$.

\begin{figure}
\centerline{\epsfxsize=\columnwidth \epsfbox{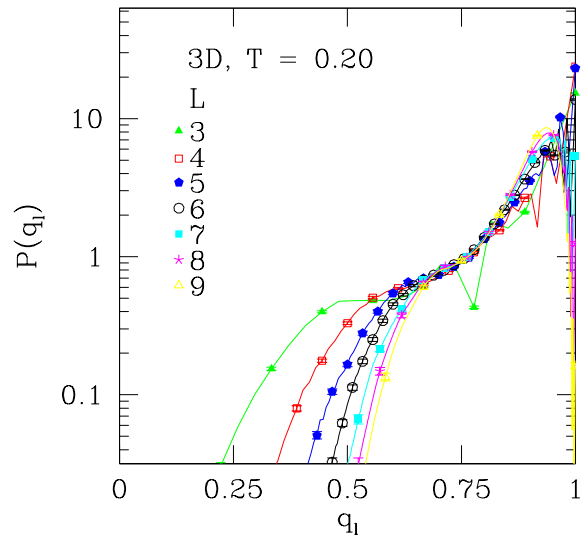}}
\caption{
The distribution of the link overlap in 3D at $T=0.20$ with
free boundary conditions for
different sizes. Note the logarithmic vertical scale.
}
\label{pqb0.20_3d-fbc}
\end{figure}

\begin{figure}
\centerline{\epsfxsize=\columnwidth \epsfbox{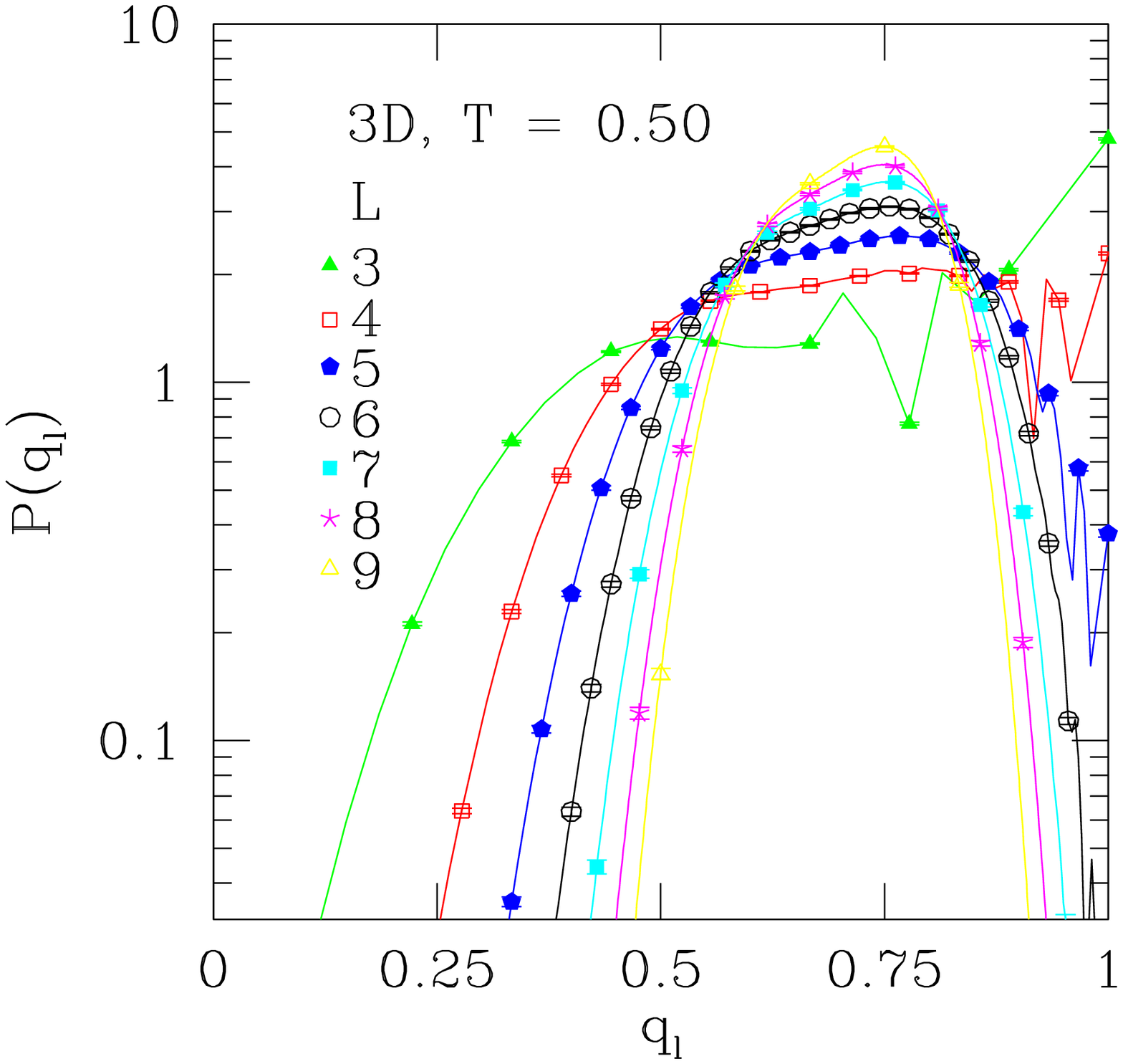}}
\caption{
Same as for Fig.~\ref{pqb0.20_3d-fbc} but at $T = 0.50$.
}
\label{pqb0.50_3d-fbc}
\end{figure}

Figures \ref{pqb0.20_3d-fbc} and \ref{pqb0.50_3d-fbc} show data for the link
overlap $\ql$ at $T=0.20$ and $0.50$, respectively. We see a large peak for 
large $\ql$ values. The data for $T=0.20$ show a small shoulder for smaller 
$\ql$ values. This feature has been observed
in Ref.~\onlinecite{katzgraber:01}.

\begin{figure}
\centerline{\epsfxsize=\columnwidth \epsfbox{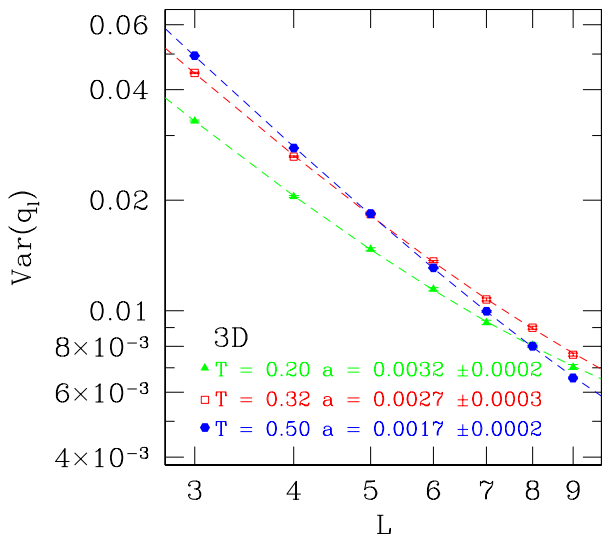}}
\caption{ Log-log plot of the variance
of $\ql$ as a function
of size in 3D at several temperatures. The dashed lines correspond to fits of
the form $y_1 = a + bL^{-c}$. 
}
\label{varqb_3d-fbc}
\end{figure}

Figure \ref{varqb_3d-fbc} shows data for the variance of $P(\ql)$ at several
temperatures. We attempt two fits of the form
\be
\varql = a + bL^{-c}
\label{fitfunc.1}
\ee
and
\be
\varql = dL^{-e} \; .
\label{fitfunc.2}
\ee

Table \ref{varql-3d-fbc.1} shows the relevant parameters for the
three-parameter fits to
Eq.~(\ref{fitfunc.1}).
From our data, we see a finite value of $a$ for the sizes
studied, a result that implies $d = d_s$, as expected in RSB.
A two-parameter fit of the form $\varql \sim L^{-\mu_l}$ in
Eq.~(\ref{fitfunc.2}) does not seem plausible
given the small fitting probabilities $Q$ 
of $10^{-18}$, $10^{-37}$, and $0.0$ for $T = 0.20$, $0.34$, and $0.50$,
respectively. Similar results are found for 
zero-temperature calculations \cite{palassini:01a}. 

However, given the modest range of sizes studied, and the likelihood that there are strong finite-size corrections with free boundary conditions,
we cannot rule out that asymptotically one has
$a = 0$, implying $d_s < d$.
Evidence for large finite-size corrections with free boundary conditions comes
from 
our estimates for $T_c$, which are
inconsistent with known results\cite{marinari:98} 
by $\sim 10\%$.

\begin{table}
\caption{
\label{varql-3d-fbc.1}
Fits for $\varql$ for the three-dimensional EA Ising spin-glass
with free boundary conditions to the form in Eq.~(\ref{fitfunc.1}).
We show the data for different temperatures and find a finite value for $a$.
The fit probabilities $Q$ are reasonable for the lower
temperatures.
}
\begin{tabular*}{\columnwidth}{@{\extracolsep{\fill}} c r r r l }
\hline 
\hline
$T$  &  $a$  & $b$ & $c$ & $Q$ \\ 
\hline
0.20 & $0.0032 \pm 0.0002$ & $0.228 \pm 0.006$ & $1.860 \pm 0.025$ & $0.819$ \\
0.34 & $0.0027 \pm 0.0003$ & $0.349 \pm 0.014$ & $1.937 \pm 0.038$ & $0.003$ \\
0.50 & $0.0017 \pm 0.0002$ & $0.463 \pm 0.014$ & $2.072 \pm 0.029$ & $0.000011$ \\
\hline
\hline
\end{tabular*}
\end{table}

\subsection{Four dimensions}
\label{results.4d}

Table \ref{4d-tab} shows the parameters of the simulation in four dimensions. 
Our lowest simulated temperature is $0.20$ (to be compared with
$T_c \approx 1.80$)\cite{parisi:96} and the largest is $2.80$. 
We use the same temperature set for each size studied. The acceptance ratios
for the parallel tempering method are typically greater than $0.3$.

\begin{table}
\caption{
\label{4d-tab}
Parameters of the simulations in four dimensions
with free boundary conditions.
}
\begin{tabular*}{\columnwidth}{@{\extracolsep{\fill}} c r r l }
\hline  
\hline
$L$  &  $\nsa$  & $\nsw$ & $N_T$  \\ 
\hline
3 & 10000 & $6 \times 10^3$ & 23 \\
4 & 10000 & $6 \times 10^4$ & 23 \\
5 &  5000 & $5 \times 10^5$ & 23 \\
\hline
\hline
\end{tabular*}
\end{table}

\begin{figure}
\centerline{\epsfxsize=\columnwidth \epsfbox{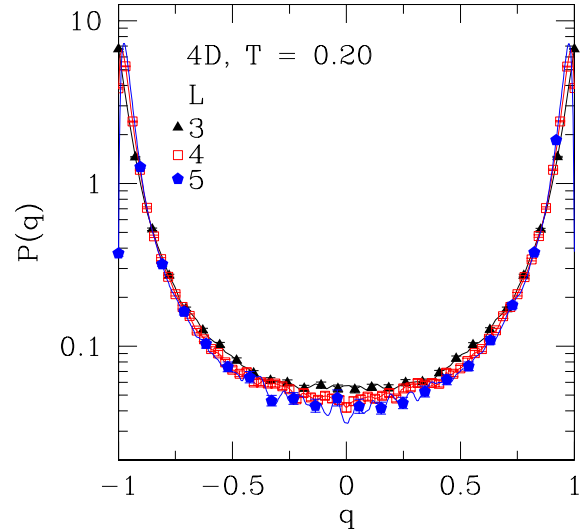}}
\caption{
Data for the overlap distribution $P(q)$ in 4D at $T=0.20$ with free boundary
conditions.  The data are normalized, so the area under the curve is unity.
}
\label{pq0.20_4d-fbc}
\end{figure}

\begin{figure}
\centerline{\epsfxsize=\columnwidth \epsfbox{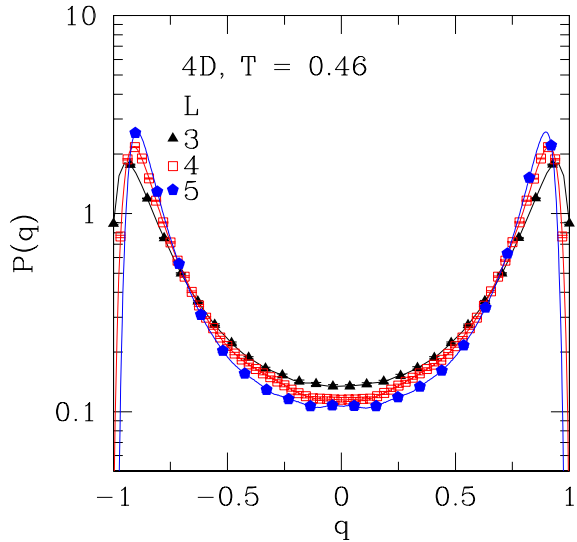}}
\caption{Same as for Fig.~\ref{pq0.20_4d-fbc} but at $T=0.46$.
}
\label{pq0.46_4d-fbc}
\end{figure}

\begin{figure}
\centerline{\epsfxsize=\columnwidth \epsfbox{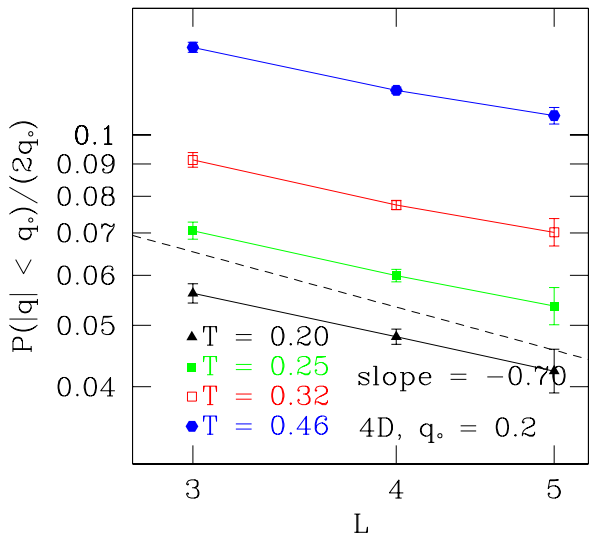}}
\caption{ Log-log plot of $P(0)$ against $L$ in 4D
averaged over the range $|q| < q_\circ = 0.20$. For small sizes a decrease in
$P(0)$
with increasing system size is visible.
}
\label{p0_4d-fbc}
\end{figure}

Figures \ref{pq0.20_4d-fbc} and \ref{pq0.46_4d-fbc} show data for $P(q)$ at
temperatures 0.20 and 0.46. As in three dimensions, the tail of the
distribution depends on the system size $L$. To better quantify this behavior,
we show in Fig.~\ref{p0_4d-fbc} data for $P(0)$ vs $L$ averaged over the
range $|q|< 0.20$.  Since the DP predicts that $P(0) \sim L^{-\theta}$,
where\cite{hartmann:99a,hukushima:99} $\theta \simeq 0.7$ we also indicate 
this behavior by the dashed line in the figure. The data is reasonably
consistent with this.
Fixing 
$\theta^\prime = 0.70$
and performing a fit of the form $a L^{-\theta^\prime}$,
we find goodness-of-fit probabilities $Q$ of $0.53$, $0.44$, and $0.21$ for 
$T = 0.20$, $0.25$, and $0.32$, respectively. Fixing $\theta^\prime = 0$ gives
us goodness-of-fit probabilities $Q< 10^{-4}$.
Unfortunately, our modest range of sizes does not permit us to make more
quantitative statements and a crossover to a behavior where
$P(0) \sim L^{0}$ cannot be excluded.
As in three dimensions, our data is consistent with
$P(0) \propto T$ for $T \rightarrow 0$.

\begin{figure}
\centerline{\epsfxsize=\columnwidth \epsfbox{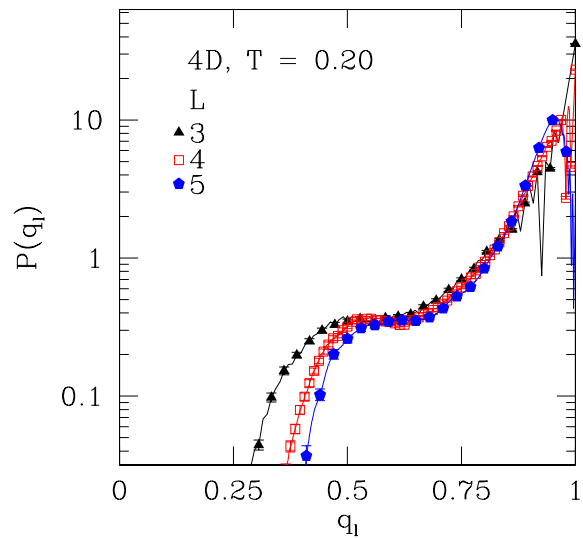}}
\caption{ The
distribution of the link overlap in 4D at $T=0.20$ for different
sizes with free boundary conditions. Note the logarithmic vertical scale. }
\label{pqb0.20_4d-fbc}
\end{figure}

\begin{figure}
\centerline{\epsfxsize=\columnwidth \epsfbox{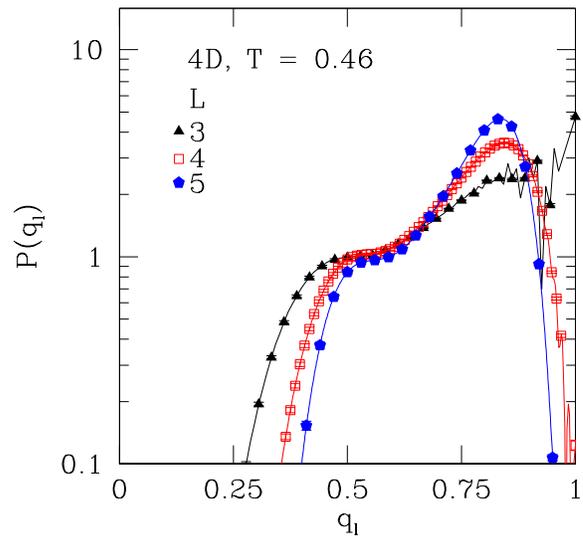}}
\caption{
Same as for Fig.~\ref{pqb0.20_4d-fbc} but at temperature 0.46.
}
\label{pqb0.46_4d-fbc}
\end{figure}

The data for the distribution of the link overlap is shown in
Figs.~\ref{pqb0.20_4d-fbc} and \ref{pqb0.46_4d-fbc} for temperatures
0.20 and 0.46, respectively. As in three dimensions, we see a double-peak
structure.

\begin{figure}
\centerline{\epsfxsize=\columnwidth \epsfbox{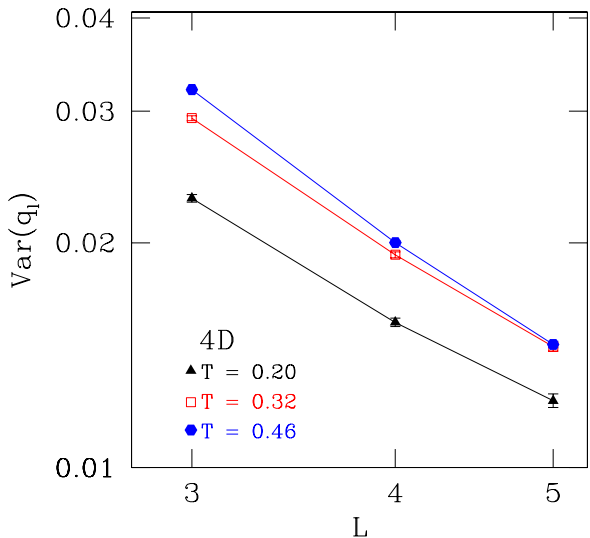}}
\caption{ Log-log plot of the variance 
of $\ql$ as a function
of size in 4D at several temperatures with free boundary conditions. 
The data do not fit well to a power law.
}
\label{varqb_4d-fbc}
\end{figure}

In Fig.~\ref{varqb_4d-fbc} we show the variance of the link overlap $\ql$ as a
function of system size for several low temperatures. We perform fits to the
functions indicated in Eqs.~(\ref{fitfunc.1}) and (\ref{fitfunc.2}). 
Table \ref{fitres.4d.1} shows the
results of the three-parameter fits to Eq.~(\ref{fitfunc.1}).
The best fit has
$a > 0$, a result compatible with RSB, but, because we have the same number of
data points as parameters, we cannot give an error bar or fit probability $Q$.
Table \ref{fitres.4d.2} shows the
results of the two-parameter fits to Eq.~(\ref{fitfunc.2}).
The goodness-of-fit probabilities are quite small, but not impossibly so for
small $T$, so we cannot rule out
a scenario in which $\varql \to 0$ for $L \to \infty$.

\begin{table}
\caption{
\label{fitres.4d.1}
Fits for $\varql$ for the four-dimensional EA Ising spin glass
with free boundary conditions for the fit in Eq.~(\ref{fitfunc.1}).
We show the data for different temperatures and find a finite value for $a$.
We cannot quote error bars or fitting
probabilities since we have the same number of data points as variables.
}
\begin{tabular*}{\columnwidth}{@{\extracolsep{\fill}} c r r l }
\hline 
\hline
$T$  &  $a$  & $b$ & $c$ \\
\hline
0.20 & $0.0063$ & $0.1517$ & $2.0156$ \\
0.32 & $0.0057$ & $0.1980$ & $1.9342$ \\
0.46 & $0.0057$ & $0.2729$ & $2.1285$ \\
\hline
\hline
\end{tabular*}
\end{table}

\begin{table}
\caption{
\label{fitres.4d.2}
Fits for $\varql$ for the four-dimensional EA Ising spin glass
with free boundary conditions for the fit in Eq.~(\ref{fitfunc.2}).
The probabilities for the fit are quite small but not impossibly so
for small $T$.
}
\begin{tabular*}{\columnwidth}{@{\extracolsep{\fill}} c r r l }
\hline   
\hline
$T$  &  $d$  & $e$ & $Q$ \\
\hline
0.20 & $0.090 \pm 0.008$ & $1.255 \pm 0.005$ & $0.09$ \\
0.32 & $0.137 \pm 0.010$ & $1.406 \pm 0.003$ & $0.05$ \\
0.46 & $0.179 \pm 0.015$ & $1.572 \pm 0.002$ & $0.0008$ \\
\hline
\hline
\end{tabular*}
\end{table}

\section{Conclusions}
\label{conclusions}

We have performed Monte Carlo simulations of the three- and
four-dimensional EA Ising spin glass with free boundary conditions.
For small sizes, we find $P(0) \sim L^{-\theta^\prime}$
with $\theta^\prime > 0$.
The values of $\theta^\prime$ we obtain for the lowest temperatures simulated
are compatible with predictions from the droplet model. 
However, in three dimensions,
we find evidence for crossover to a
behavior where $P(0) \sim L^{0}$ for larger
system sizes. In four dimensions, where
the range of sizes is smaller, we do not see
evidence for this crossover.
An analysis of our results indicates that the surface of these excitations is
space filling, corresponding to RSB.
However, since free boundary
conditions have large finite-size corrections, it is not clear if these
results
represent the asymptotic behavior. Overall, our results are quite similar to
those of Ref.~\onlinecite{palassini:01a}, which were
obtained at zero temperature.

Although free boundary conditions have large finite-size corrections because a
substantial fraction of spins are on the surface, there is also a
compensating benefit 
in that they do not pose any restrictions on the
position of domain walls (see Fig.~\ref{fbc.sketch}). It would be interesting
to
investigate what the optimal boundary conditions for spin-glass studies.

\begin{acknowledgments}
We would like to thank M.~Palassini for useful discussions.
This work was supported by the National Science Foundation under grant 
No.~DMR 0086287.
The numerical calculations were made possible by use of the UCSC 
Physics graduate computing cluster funded by the Department of
Education Graduate Assistance in the Areas of National Need program.
We also thank Gary Glatzmaier for allowing us to use the
32-node Beowulf cluster at the UCSC Earth Sciences Department, which is
funded by
NASA's Planetary Geology and Geophysics Program. 
\end{acknowledgments}

\bibliography{refs}

\end{document}